\documentclass[prl, twocolumn, showpacs,preprintnumbers,amsmath,amssymb,tightenlines,epsfig,floatfix]{revtex4}

\usepackage{graphicx}
\usepackage{dcolumn}
\usepackage{bm}
\usepackage{amsfonts}

\begin{document}
\preprint{Version: submission \today}
\title{Generating squeezing in an atom laser through self-interaction}
\author{Mattias T. Johnsson and Simon A. Haine}
\affiliation{Australian Centre for Quantum Atom Optics, The Australian National University, Canberra, 0200, Australia.}

\begin{abstract}
We describe a scheme for creating quadrature- and intensity-squeezed atom lasers that do not require squeezed light as an input. The beam becomes squeezed due to nonlinear interactions between the atoms in the beam in an analogue to optical Kerr squeezing. We develop an analytic model of the process which we compare to a detailed stochastic simulation of the system using phase space methods. Finally we show that significant squeezing can be obtained in an experimentally realistic system and suggest ways of increasing the tunability of the squeezing. 
\end{abstract}

\pacs{03.75.Pp,  42.50.Dv, 03.75.Gg}                                                                                                                                       

\maketitle
{\it Introduction.---} The creation of the optical laser and the development of quantum optics has allowed tests of many fundamental properties of quantum mechanics \cite{aspect82, kimble92, furusawa98}. The ability to create quadrature squeezing is an important prerequisite for many of these tests as it allows the creation of continuous variable entanglement between the amplitude and phase of two spatially separated optical beams \cite{kimble92,bowen_bias_entangle}. 
 With the advent of the atom laser, there is much interest in creating a quadrature-squeezed atomic beam as it allows us to revisit many of these tests using massive particles rather than photons. For example, it has been shown that continuous variable entanglement between the amplitude and phase of spatially separated atomic beams for use in Einstein-Podolsky-Rosen (EPR) tests \cite{reid} can be generated by dissociation of a molecular Bose-Einstein condensate (BEC) \cite{kheruntsyan2005}, or by outcoupling from a BEC using a Raman transition with squeezed light \cite{haine05}.
Another example is interferometry --- the use of massive particles over photons already offers the promise of vastly improved sensitivity \cite{petersET1997}, and quadrature squeezing offers the possibility of going beyond this to beat the standard quantum shot-noise limit \cite{dowling1998}.

The standard scheme to create a squeezed atom laser is to use a squeezed optical field to couple atoms out of the BEC and into the atom laser beam, attempting to transfer the quantum state of the light onto the atoms \cite{jingET2000,fleischhauerET2002,haineET2006}. Such a scheme is challenging, as it requires squeezed light at the relevant transition frequencies of the atomic species making up the BEC. Obtaining useful amounts of squeezing at these frequencies is a hard problem, although recently there has been some success \cite{riesET2003,tanimuraET2006,mccormickET2006,hetetET2007}.

In this Letter we describe a scheme to generate a quadrature-squeezed atom laser without a squeezed optical field, thus removing a significant source of complexity, and model the effect for an experimentally realistic system. Our scheme utilizes the nonlinear interaction caused by atom-atom scattering to create a Kerr squeezing effect \cite{wallsET1994}.
The rate at which the beam squeezes is dependent purely on the local density of the beam. As the atoms fall under gravity, the beam become more dilute, leading to a continuous reduction in the strength of the Kerr effect along the length of the beam. This ensures the Kerr effect acts only for a finite time, preventing the squeezing from becoming degraded in the long term limit.  

The possibility of nonlinearities resulting in quadrature squeezing has previously been considered by Jing \textit{et al.} using a zero-dimensional, single-mode analysis \cite{jingET2001}, who found very little squeezing within the range of validity of their linearized model. It has not been considered for a realistic atom laser, taking into account multimode effects, non-Markovian behavior (i.e. back coupling of the beam into the BEC) and mode matching.

The structure of this Letter is as follows: We develop a single-mode model to obtain an analytic expression for the squeezing. We then create a realistic model of a Raman atom laser and simulate it using a stochastic phase space approach. These simulations are compared to a spatially integrated version of the analytic solution to determine its predictive power. Finally we consider the implications for current experimental atom lasers.

{\it Analytic model.---} We first construct a single-mode model of Kerr squeezing that admits an analytic solution. We use the Kerr Hamiltonian
\begin{equation}
\hat{H} = \hbar \omega  \hat{a}^{\dagger} \hat{a} +\frac{ \chi}{2}  \hat{a}^{\dagger} \hat{a}^{\dagger} \hat{a}  \hat{a}
\label{eqSingleModeHamiltonian}
\end{equation}
where $\chi$ is the strength of the nonlinearity and $\hat{a}$
describes a bosonic field. In the Fock basis, the evolution of a system governed by (\ref{eqSingleModeHamiltonian}) is described by $| \psi(t) \rangle = \sum_n c_n(t) | n \rangle $, with $c_n(t) = c_n(0)\exp[-i(n \omega +\chi n(n-1)/2 \hbar)t]$.

\begin{figure}[htb]
\begin{center}
\includegraphics[width=7.5cm,height=5.5cm]{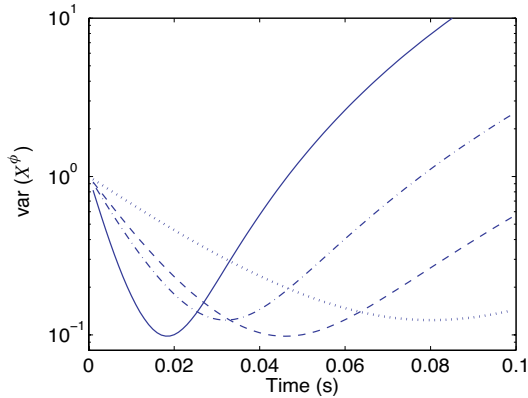}
\caption{Single-mode Kerr squeezing as a function of time for a nonlinearity $\chi$ and particle number $N=\alpha^2$. $\chi=0.1\hbar$, $\alpha = \sqrt{1000}$ (solid); $\chi=0.04\hbar$, $\alpha = \sqrt{1000}$ (dashed);  $\chi=0.1\hbar$, $\alpha = \sqrt{500}$ (dash-dotted);  $\chi=0.04\hbar$, $\alpha = \sqrt{500}$ (dotted).  } 
\label{figAnalyticSqueezing}
\end{center}
\end{figure}

In order to examine squeezing in this system we define the standard quadrature operator $\hat{X}^{\phi} = e^{i\phi} \hat{a} + e^{-i\phi} \hat{a}^{\dagger}$,
where $\phi$ is the phase angle at which the measurement is carried out. The variances of the $\hat{X}^{\phi}$ for all $\phi$ are unity for a coherent state, and consequently a state is squeezed if the variance is less than one for a particular value of $\phi$. If we assume that our system is initially in a coherent state $|\psi(0)\rangle = |\alpha\rangle$, then defining ${\mathrm{var}}(X^{\phi}) = \langle \psi(t)| \hat{X}^{\phi2} |\psi(t) \rangle - \langle \psi(t)| \hat{X}^{\phi} |\psi(t) \rangle^2$ gives
\begin{eqnarray}
{\mathrm{var}}(X^{\phi}(t)) &=&  1 + 2\alpha^2   \nonumber \\
&&\hspace{-2cm} + 2\alpha^2 \exp\left[-2\alpha^2 \sin^2\frac{\chi t}{\hbar}\right]  \cos \left[ \frac{\chi t}{\hbar} +\alpha^2 \sin \frac{2 \chi t}{\hbar} -2\phi \right] \nonumber \\
&& \hspace{-2cm}  - 4\alpha^2 \exp \left[ -4\alpha^2 \sin ^2 \frac{\chi t/2}{\hbar}  \right] \cos^2 \left[ \phi - \alpha^2 \sin\frac{\chi t}{\hbar} \right]  \label{eqVarianceOfX1SingleMode}.
\end{eqnarray}
At any given time there is an optimum choice of $\phi$ that 
gives the best squeezing. Plots of the minimum value of ${\mathrm{var} (\hat{X}^{\phi})}$ over time for a variety of nonlinear interaction strengths are shown in Fig.\ \ref{figAnalyticSqueezing}, where we have chosen $\phi$ to give the lowest possible variance of $\hat{X}^{\phi}$. The time taken to reach best squeezing scales inversely with $\chi$ while the minimum variance is given by $\sim \alpha^{-2/3}$ for $\alpha>3$.
In this model arbitrarily good squeezing can be obtained provided the number of particles in the system can be arbitrarily large.

{\it Stochastic simulation of an atom laser.---} We now develop a realistic, multimode and spatially extended description of an atom laser. A mean field analysis using the Gross Pitaevskii equation (GPE) will not be adequate, as the GPE is incapable of examining the quantum statistics of the system, and thus cannot say anything about squeezing. We therefore model the system using stochastic phase space methods \cite{gardinerET2000}.  This involves finding the master equation for the system and then converting to a specific representation of the probability distribution to obtain a Fokker-Planck equation (FPE). This equation can then be treated as a set of stochastic partial differential equations (PDEs) which can be solved numerically.

Our model is based on a Raman atom laser \cite{edwards99, hagleyET1999, robinsET2006}. After adiabatically eliminating the excited state, the effective Hamiltonian for the system is
\begin{eqnarray}
\hat{H}_{\mathrm{eff}} &=& \int d{\mathbf{r}} \left[ \hat{\Psi}^{\dagger}_1 \left( -\frac{\hbar^2}{2m}\nabla^2 + \frac{1}{2} m \omega^2 r^2 - \frac{\hbar |\Omega_{13}|^2} {\Delta_{13}} \right) \hat{\Psi}_1 \right. \nonumber \\
&& \left. + \hat{\Psi}^{\dagger}_2 \left( -\frac{
\hbar^2}{2m}\nabla^2 -   \frac{\hbar}{\Delta_{13}} |\Omega_{23}|^2 - \hbar \delta \right)  \hat{\Psi}_2\right. \nonumber \\
&& + \frac{1}{2} \sum_{i=1}^2 U_{ii} \hat{\Psi}^{\dagger}_i \hat{\Psi}^{\dagger}_i \hat{\Psi}_i \hat{\Psi}_i  + U_{12}  \hat{\Psi}^{\dagger}_1 \hat{\Psi}^{\dagger}_2 \hat{\Psi}_2 \hat{\Psi}_1 \nonumber \\
&&  \left. -\frac{\hbar}{\Delta_{13}} \left( \Omega_{13} \Omega_{23}^* e^{- i({\mathbf{k}}_0 \cdot {\mathbf{r}} - \delta t)} \hat{\Psi}_2^{\dagger} \hat{\Psi}_1 + {\mathrm{H.c.}} \right) \right]
\label{eqHeff} 
\end{eqnarray}
where $\hat{\Psi}_1({\mathbf{r}})$ and $\hat{\Psi}_2({\mathbf{r}})$ describe the trapped and untrapped matter fields respectively, $\Delta_{1j}$ the single-photon detunings, $\delta$ the two-photon detuning, $\Omega_{13}$ and $\Omega_{23}$ the Rabi frequencies of the two optical fields, ${\mathbf{k}}_0 = {\mathbf{k}}_2 - {\mathbf{k}}_1$ the momentum kick imparted to the outcoupled atoms (taken to be downward in our simulations), $\omega$ the harmonic trap frequency and $U_{ij} = 4 \pi \hbar^2 a_{ij}/m$, where $a_{ij}$ is the $s$-wave scattering length between atoms in states $|\Psi_i\rangle$ and $|\Psi_j\rangle$. As the matter fields are position dependent, Eq.\ (\ref{eqHeff}) describes the full multimode nature of the problem including non-Markovian effects. 

We will work in the Wigner representation, but ignore third and higher order derivatives in the FPE as these terms do not have a simple mapping to stochastic PDEs, and can be assumed to be negligible when the field has a high occupation number \cite{norrieET2006}. This truncated Wigner approximation (TWA) will eventually fail, but over the timescale of our simulations the TWA was indistinguishable from the exact analytic solutions we had in the single mode case. In addition, quantities accessible by multimode GPE simulations also agreed with the TWA solutions over these timescales. The stochastic PDEs describing the system in the TWA are
\begin{eqnarray}
i \hbar \frac{\partial \psi_1}{\partial t} &=& \left( \frac{-\hbar^2}{2m} \nabla^2 + \frac{1}{2} m \omega^2 r^2 - \frac {\hbar |\Omega_{13}|^2} {\Delta_{13}} \right. \nonumber \\
&& + U_{11}(|\psi_1|^2 - \frac{1}{\Delta V})  + U_{12} (|\psi_2|^2 - \frac{1}{2\Delta V}) \bigg) \psi_1 \nonumber \\
&& - \hbar \Omega e^{i {\mathbf{k}}_0 \cdot {\mathbf{r}} } \psi_2  \label{eqStochPDEpsitrapped} \\
i \hbar \frac{\partial \psi_2}{\partial t} &=& \left( \frac{-\hbar^2}{2m} \nabla^2 - \frac{\hbar |\Omega_{23}|^2} {\Delta_{13}} - \hbar \delta + U_{22}(|\psi_2|^2 -\frac{1}{\Delta V}) \right. \nonumber \\
&& \left. + U_{12} (|\psi_1|^2 -\frac{1}{2\Delta V})  \right) \psi_2 - \hbar \Omega^* e^{-i {\mathbf{k}}_0 \cdot {\mathbf{r}}} \psi_1 \label{eqStochPDEpsiuntrapped}
\end{eqnarray}
where $ \Omega = \Omega_{13}^* \Omega_{12}/\Delta_{13}$ is the two-photon Raman Rabi frequency, $\psi_1$ and $\psi_2$ are the c-number stochastic variables corresponding to the quantum operators for the trapped matter field and the atom laser beam respectively, and $\Delta V$ is the spacing of the grid on which the problem is to be numerically simulated. The terms inversely proportional to $\Delta V$ compensate for the mean field of the vacuum, which is non-zero in the Wigner approach. As the FPE has no second order derivative term, there are no explicit noise terms in the equations. Noise still enters the problem, however, as we must include the correct noise in the initial conditions for Eqs.\ (\ref{eqStochPDEpsitrapped}) and (\ref{eqStochPDEpsiuntrapped}). We chose this initial noise such that it corresponded to a coherent state. In all simulations parameters appropriate to a Rb Raman atom laser were chosen, i.e. $a=5.77$nm, $m=1.44\times 10^{-25}\,$kg, $k_0 = 2\times 10^{7}\,$m$^{-1}$, $\Omega = 50\,$rad$\,$s$^{-1}$, $\omega = 80\,$rad$\,$s$^{-1}$ and a condensate with $5\times 10^{5}$ atoms. As the simulations were carried out in one dimension, we assumed a cross-sectional area of $1.2\times 10^{-11}\,$m$^2$, and scaled $U_{ij}$ accordingly. The BEC nonlinearity $U_{11}$ was set to zero; this restriction will be discussed later. The stochastic equations (\ref{eqStochPDEpsitrapped}) and (\ref{eqStochPDEpsiuntrapped}) were solved numerically using the open source package XMDS \cite{XMDS}.

\begin{figure}[htb]
\begin{center}
\includegraphics[width=7cm,height=5cm]{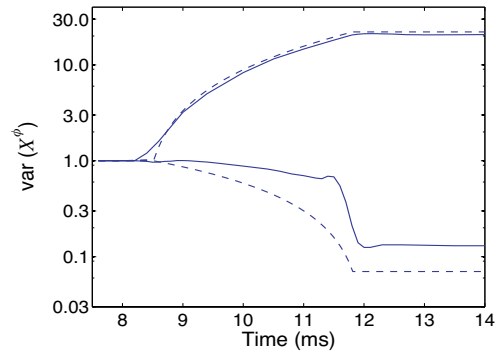}
\caption{Variance of the quadratures $\hat{X}^{\phi}$ (lower trace) and $\hat{X}^{\phi + \pi/2}$ (upper trace) as a function of time for the atom laser beam in a region $20\, \mu$m long well below the condensate. The beam reaches the region $9\,$ms after outcoupling begins. The solid traces represent the results from the multimode stochastic simulation, and dashed traces represent our single-mode analytic model.  After $12\,$ms steady state is reached in the region of interest.} 
\label{figAnalyticVsStochasticComparison_new}
\end{center}
\end{figure}

Unlike the single-mode, zero-dimensional analytic model discussed earlier, the beam of an atom laser is an extended object, so we cannot talk of a particle number in the beam given by $N=|\alpha|^2$ as we could in the analytic case. In a beam the relevant quantity is the local density $\rho$. To quantify the squeezing in the multimode case, we define the amplitude and phase quadrature operators by picking a particular spatial mode of the quantum field. We define the amplitude and phase quadrature operators as $\hat{X}^{\phi} = \hat{b} + \hat{b}^{\dag}$, where $\hat{b} = \int_{z_1}^{z_2} L^{*}(z)\hat{\psi}_2(z)\, dz$. Here, $L(z)$ represents the spatial mode in which we are interested, and $\int_{z_1}^{z_2}|L(z)|^2 dz =1$. Physically, the form of $L(z)$ would be determined by the form of the local oscillator used in a homodyne measurement. The details of how the squeezing would be measured in an experiment will be considered in the Discussion section. Maximum squeezing will be observed when $L(z)$ best matches the spatial mode of the field in which the squeezing occurs. We chose $L(z)$ as a plane wave with wavelength and frequency that best matched the atomic beam over the region $z_1 \rightarrow z_2$,  i.e. $L(z) = e^{i(k_L z - \omega_L t +\phi)}$, with $k_L = k_0 - U_{22}\rho m/k_0 \hbar^2$, and $\omega_L = \hbar k_L^2/2m + U_{22}\rho/\hbar$. In Fig.\ \ref{figAnalyticVsStochasticComparison_new} we plot the results of a stochastic simulation showing the variance of (lower solid trace) $\hat{X}^{\phi}$ and $\hat{X}^{\phi+ \pi/2}$ (upper solid trace). $\phi$ was chosen to minimise the variance of $\hat{X}^{\phi}$ (the squeezing quadrature), and hence maximise the variance of $\hat{X}^{\phi + \pi/2}$ (the anti-squeezing quadrature). Initially the quadrature variance is unity, as only vacuum is present. As the beam traverses the region, the overlap between the atom laser beam and $L(z)$ becomes high, and squeezing is measured, reaching steady state shortly after the beam front has completely passed through the region.
As the system reaches steady state mode matching is achieved, resulting in the reduced variance signifying squeezing in the mode $L(z)$.  

We now consider to what extent our simple analytic model correctly predicts the squeezing. To compare the analytic model with the multimode simulations we choose $\chi = U_{22}\int |L(z)|^4 \, dz$, $\alpha = \sqrt{N}$ where $N = \int_{z_1} ^{z_2} |L(z)|^2 \left( |\psi_2(z)|^2 - 1/2\Delta V \right) dz$, and then average the var$(X^\phi)$ predicted by the analytic model over a period of time which corresponds to the time required for atoms to pass through the region $[z_1, z_2]$. The results are also shown in Fig.\ \ref{figAnalyticVsStochasticComparison_new}, which compares the best squeezing and antisqueezing predicted by the single-mode analytic model to the results of the stochastic simulations. 
The antisqueezing is very well predicted, but the single-mode model predicts squeezing almost two times better than is actually seen. This discrepancy is largely due to the difficulty of mode matching the multimode beam to the local oscillator. Any mode matching discrepancy will have a larger relative effect on the measured squeezing than the antisqueezing.

{\it Discussion.---} Although our numerical simulations were carried out in 1D, our scheme still functions for a real, 3D system. The main difference between 1D and 3D is that 3D allows the beam to have a transverse mode structure. Theoretically this structure is irrelevant as the squeezing rate is purely dependent on the particle density per mode, independent of dimension, and the atom laser tends to single mode operation in the long time limit. It is still desirable to have as little transverse structure as possible, however, as this reduces mode matching problems. We now consider a realistic, 3D example. A Rb Raman atom laser, such as the one described in \cite{robinsET2006}, with mean trapping frequency $\bar{\omega} = 2\pi(60\times 600 \times 600)^{1/3}$, has an atomic density just below the condensate of $\rho_0 \sim 3\times 10^{18}\,$m$^{-3}$ if a two-photon Rabi frequency $\Omega=500\,$rad$\,$s$^{-1}$ is chosen. A Raman atom laser is minimally divergent \cite{robinsET2006}, so density scales only due to acceleration by gravity. After falling a distance $z$ the beam density is $\rho = \rho_0 /(1+ m\sqrt{2 g z}/\hbar k_0)$. Assuming $k_0 = 3.2\times 10^7\,$m$^{-1}$ and that the mode match region is a section of the beam 25$\mu$m in vertical extent $1\,$cm below the condensate, there are $\sim 1100$ atoms in this region. Using these numbers our integrated analytic model predicts ${\mathrm{var}}(\hat{X}^\phi) = 0.143$, and ${\mathrm{var}}(\hat{X}^{\phi+\pi/2}) = 7.11$, where we use a time-dependent $\chi$ to model the density decrease as the atoms fall. Using Bragg diffraction as a beam splitter, squeezing of this level leads to entanglement under the Reid-Drummond criterion \cite{reid}. While the measured amount of squeezing will not reach this due to mode matching difficulties, it indicates our scheme is certainly feasible. If the nonlinearity or the density of the beam could be increased, the squeezing would further increase. Possible mechanisms to accomplish this might be the use of Feshbach resonances to increase the nonlinearity \cite{robertsET1998}, or the use of far-detuned light fields to focus the atom laser beam and increase the atomic density.
 
The flexibility of this scheme is clear: As the best squeezing depends only on the local density of the beam and how long atoms have been in the beam when they are measured, and since it is possible to tune the outcoupling strength, momentum kick and place of measurement, there is a large parameter regime over which good squeezing can be obtained.
 
Our scheme relies on the output beam starting in a coherent state. As the outcoupling process functions as a beam splitter some of the quantum statistics of the BEC will be copied onto the beam. Assuming the BEC itself begins in a coherent state, it will also exhibit Kerr squeezing. However, due to the BEC's much higher density, it will reach peak squeezing more quickly than the beam, after which the squeezing will degrade as the nonlinearities cause its phase to become uncertain. The long-time limit of such a process is var$(\hat{X}^\phi _{bec}) = 2N$ for any quadrature, where $N$ is the BEC particle number. As the outcoupling is weak, the beam will only weakly reflect the quantum statistics of the BEC, but due to the high variances this could still degrade the squeezing of the beam.

There are at least two ways to obviate this problem. The first is to reduce the nonlinearities in the condensate using a technique such as Feshbach resonances. The nonlinearities need to be suppressed such that the minimum shown in Fig.\ \ref{figAnalyticSqueezing} occurs at time comparable to the length of the experiment, meaning the suppression factor can easily be extracted from Eq.\ (\ref{eqVarianceOfX1SingleMode}). For example, in the case of the Rb laser described earlier, the BEC nonlinearity need be reduced by a factor of approximately eight hundred. The second approach is to ensure the condensate remains near a coherent state due to continuous measurement and quantum back action. For example, one could use the scheme described in Ref.\ \cite{thomsenET2002}, where a weak light beam continuously measures the condensate density. 

A homodyne measurement is required in order to observe quadrature squeezing \cite{bachor_book}. We note that this may prove challenging as obtaining a strong local oscillator which does not itself undergo Kerr squeezing may be difficult. A similar problem exists in detecting optical Kerr squeezing generated in nonlinear fibers, and can be solved by using an asymmetric Sagnac interferometer to slightly rotate the quadrature axis of best squeezing so that it lies along the amplitude quadrature \cite{schmittET1998}. Now no local oscillator is necessary as the squeezing appears in the beam intensity, and can be measured simply by performing measurements of the flux. The analogy for an atomic beam involves mixing the squeezed beam with a weak reference beam sourced from the same BEC, with the phase chosen to rotate the axis of best squeezing onto the amplitude quadrature. This could be achieved by outcoupling atom laser beams in two separate internal states (for example, the $F=1$, $m_F = 0$, and $F=2$, $m_F=0$ ground states of $^{87}$Rb) which spatially overlap, and then recombine them using a microwave transition. The relative phase of the two field can be adjusted by altering the phase of the outcoupling field. As the fields spatially overlap, the mode-matching will be automatic.  We have performed an analytic, single-mode analysis of the interference of two Kerr-squeezed atom laser beams with different intensities derived from the same condensate. In the case where the relative intensity between the two beams was $0.5$, and using the same parameters used to generate Fig.\ \ref{figAnalyticVsStochasticComparison_new}, we found an intensity noise of $0.17$ as compared to unity for a coherent state when the two beams were interfered with the appropriate phase. The intensity squeezing obtained by this method is robust to changes in the intensity of the two beams with best squeezing obtained for beam intensity ratios between 0.25 and 0.50. This intensity squeezing is of considerable interest as one of the prime applications for atom lasers is precision interferometric measurements which are ultimately limited by shot noise.

We would like to thank Joseph Hope, Oliver Gl\"{o}ckl, Matthew Jeppesen and Hans Bachor for helpful discussions. This work was supported by the ARC and the APAC National Supercomputing Facility.

\end{document}